\documentclass[11pt]{article}
\usepackage[T1]{fontenc}
\usepackage[latin1]{inputenc}
\usepackage[english]{babel}
\usepackage{amsmath}
\usepackage{amsbsy}
\usepackage{latexsym}
\usepackage{amssymb}
\usepackage{anysize}
\usepackage{lscape}
\usepackage{subfigure}
\usepackage{caption}

\usepackage[final]{epsfig}
\usepackage{graphics}
\usepackage{amsfonts}
\usepackage{cancel}
\usepackage{latexsym}
\usepackage{amssymb}
\usepackage{url}
\usepackage{esvect}
\usepackage{graphicx,epsfig,psfrag}
\usepackage[left=1.4in, right=1.4in, top=1in, bottom=1in]{geometry}

\begin{document}
\newcommand{\eps}{{\varepsilon}}
\newcommand{\proofend}{$\Box$\bigskip}
\newcommand{\C}{{\mathbf C}}
\newcommand{\Q}{{\mathbf Q}}
\newcommand{\R}{{\mathbf R}}
\newcommand{\Z}{{\mathbf Z}}
\newcommand{\RP}{{\mathbf {RP}}}
\newcommand{\CP}{{\mathbf {CP}}}
\newcommand{\A}{{\rm Area}}
\newcommand{\Le}{{\rm Length}}
\newcommand{\be}[1]{\begin{equation}\label{#1}}
\newcommand{\ee}{\end{equation}}
\newcommand{\ba}[1]{\begin{eqnarray}\label{#1}}
\newcommand{\ea}{\end{eqnarray}}
\newcommand{\rf}[1]{(\ref{#1})}
\newcommand{\nn}{\nonumber}
\newcommand{\DD}{{\rm I\kern-3pt D}}
\newcommand{\RR}{{\rm I\kern-3pt R}}
\newcommand{\NN}{{\bf N}}
\newcommand{\CC}{{\mathbf C}}
\newtheorem{thm}{Theorem}
\newcommand{\debdem}{$\Box \,$}
\newcommand{\findem}{\,{\vrule height5pt width4pt depth0pt \hskip 2em}}



    \begin{center}
{\LARGE{Rotating saddle trap as Foucault's pendulum.}}\\
\vskip1cm
   { O.N. Kirillov $^1$, M. Levi $^2$}
\begin{itemize}
\item[1] {Helmholtz-Zentrum Dresden-Rossendorf,
P.O. Box 510119, D-01314 Dresden, Germany}
\item[2] {Department of Mathematics, Pennsylvania State University, University Park, PA 16802, USA.  }
\end{itemize}
\end{center}
\vskip 0.1 in

\begin{abstract}
One of the many surprising results found in the mechanics of rotating systems is the stabilization of a particle in a rapidly rotating planar saddle potential. Besides the counterintuitive stabilization,   an  unexpected  precessional motion is observed. In this note we show that this precession is due to a Coriolis--like force caused by the rotation of the potential. To our knowledge this is the first example where such force arises in an inertial reference frame. We also propose an idea of a simple mechanical demonstration of this effect.
\end{abstract}

\subsection*{Introduction.} According to Earnshaw's theorem an electrostatic potential cannot have stable equilibria, i.e. minima, since such potentials are harmonic functions. The theorem does not apply, however, if the potential depends on time; in fact,
 the 1989 Nobel Prize in physics was awarded to W. Paul \cite{Paul1990} for his invention of the trap for suspending charged particles in an oscillating electric field.
Paul's idea was to stabilize the saddle by ``vibrating'' the electrostatic field, by analogy with the so--called Stephenson-Kapitsa pendulum \cite{stephenson,kapitsa,landau-lifshits,Levi1998,Levi1999} in which the upside--down equilibrium is stabilized by  vibration of the pivot.
Instead of vibration, the saddle can also be stabilized by rotation of the potential (in two dimensions); this has been known for nearly a century: as early as 1918, L.E.J. Brouwer (1881--1966), one of the authors of the fixed point theorem in topology, considered stability of a heavy  particle on a rotating slippery surface, \cite{B1918,B1975,B1976}.
 \begin{figure}[thb]
	\center{  \includegraphics[angle=0, width=0.45\textwidth]{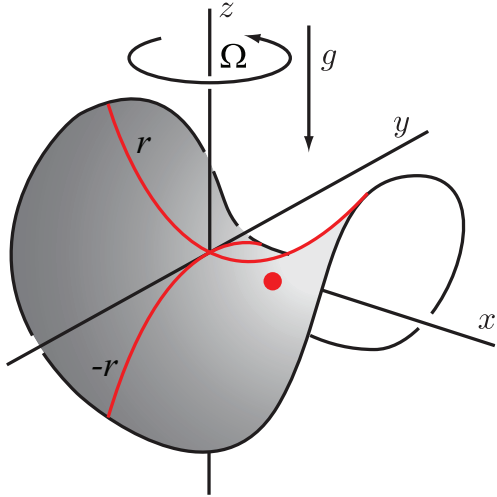}}
	\captionsetup{format=hang}
	\caption{A unit mass on a saddle \cite{B1918,B1975,B1976}.
	}
	\label{fig:figsaddle}
\end{figure}
Brouwer derived the equations of motion in  \cite{B1918,B1975};  the derivation took over 3 pages. He then linearized the equations by discarding quadratic and higher order terms in position and velocity. The resulting \textit{linearized} equations in the \textit{stationary} frame are  \cite{HB2005,S01}
\begin{equation}
     \begin{array}{l}
    \ddot x+ x \cos 2\omega t + y \sin 2\omega t=0,  \\[3pt]
    \ddot y + x \sin 2\omega t - y \cos 2\omega t =0,    \end{array}
	\label{eq:eqmotions}
\end{equation}
where $t$ is dimensionless time related to the dimensional time $\hat t$ via
$ t=  \hat t\Omega/\Omega_0 $, $\Omega_0= \sqrt{ g/r}$,\footnote{Note that $\Omega_0 $ is the frequency of small oscillations of the particle along the $x$--direction on the {\it  non--rotating} saddle. Thus our dimensionless $t$ measures time in the units of the period of the above mentioned oscillations.} and where
$ \omega $ is the dimensionless angular velocity given by  $\omega=\Omega/ \Omega_0$.

Equations   (\ref{eq:eqmotions}) look even nicer  in vector form:
\begin{equation}
	\ddot {\bf x} + S(\omega t) {\bf x} = {\bf 0},
	\label{eq:vectoreqmotions}
\end{equation}
where
\begin{equation}
	  S( \omega t)= \left( \begin{array}{rr} \cos 2 \omega t  &\  \  \sin 2 \omega t  \\
	 \sin 2 \omega t  & - \cos 2 \omega t  \end{array} \right).
		\label{eq:S}
\end{equation}
We note that these equations are written  in the stationary frame;   Brouwer actually derived the corresponding  equations in a frame rotating with the saddle (\cite{B1918,B1975,B1976}). The equations   (\ref{eq:eqmotions}), when written in the rotating frame,  acquire the Coriolis and the centrifugal forces, but lose time--dependence since the saddle appears stationary in the co--rotating frame; the equivalence of the autonomous linear equations of Brouwer and the time-periodic equations \rf{eq:eqmotions} is well-known, see e.g. \cite{S01,K2013a,K2013}.

We now describe another context in which   (\ref{eq:eqmotions})  appear.

\paragraph {A particle in a rotating potential.} The same equations   (\ref{eq:eqmotions}) -  (\ref{eq:vectoreqmotions})    arise also in a different context: they govern the motion of a unit mass in the {\it  plane} under the influence of a potential force given by the rotating saddle potential -- namely, the potential whose graph is obtained by rotating the graph of  $ z=\frac{1}{2} (x ^2 - y^2) $  around the
$z$--axis with angular velocity $\omega$.  This planar problem is related to, but is different from that of Figure~\ref{fig:figsaddle}.
Physically, such a problem arises, for instance, in the motion of a charged particle in a rotating electrostatic potential, as discussed in \cite{HB2005}.

\paragraph {Brouwer's particle vs. a particle in a rotating potential.} If    the graph of the saddle surface in Figure~\ref{fig:figsaddle} is given by a function $ z=f(x,y,t) $, then the potential energy of the unit point mass on the surface is
$ U(x,y,t) = gf(x,y,t) $. If we now  take the same function $U$ as the potential energy of a particle that lives in the {\it  plane},     we get a  problem related to Brouwer's, but not an equivalent one. To note just one aspect of the difference between the two problems, note that, unlike a particle in the plane, a particle on the surface feels centrifugal velocity--dependent forces due to the constraint to the surface. For the motions near the equilibrium these forces are quadratically small (as we will explain shortly and as was shown by Brouwer), and they disappear in the linearization. And if $U$ itself happens to be quadratic in
$x$ and $y$ (as it is in our case, as discussed   in the next section on the derivation of   (\ref{eq:eqmotions})), then the equations for the two problems are the same.

\paragraph {A particle on the surface vs. a particle in a potential.} We would like to add a general remark on the difference between two similar problems: a particle on a surface $ z=f(x,y,t) $ in a constant gravitational field on the one hand, and a particle in the plane with
a potential $ U(x,y) = g f(x,y,t)$ on the other.  The two systems have the same  potential energy -- this is the similarity.  But here is the difference:  the  kinetic energy for the particle in the planar potential is simply
 $ K_{\rm plane} =\frac{1}{2} (\dot x ^2 + \dot y  ^2 ) $, while
 \[
 2K_{\rm surface}    =   (1+f_x ^2 )\dot x ^2 + 2f_xf_y \dot x \dot y+ (1+f_y ^2 )\dot y  ^2 ;
 \]
 here $ f_x $ and $ f_y $ denote partial derivatives of the height function $f$,  is a much more complicated expression. However,  since at the  equilibrium  $ f_x= f_y=0 $,   these derivatives are small near an equilibrium, and therefore
 $ K_{\rm surface} \approx K_{\rm plane} $ there. Thus the near--equlibrium motions of the two systems are quite similar.





\paragraph {Precessional motion in the rotating saddle trap.}
It has been known since Brouwer that the motion of the particle on the saddle is stabilized for all sufficiently high $\omega$ \cite{K2013,V95}.  An illustration of a similar (but not equivalent) counterintuitive effect consists of a ball placed on a  saddle surface rotating around the vertical axis and  being in  stable equilibrium at the saddle point \cite{THB2002}. We note however that the rolling ball on a surface is a   non--holonomic system, entirely different from a particle sliding on a surface \cite{naimark-fufaev,W1997}.

\begin{figure}[h]
	\center{  \includegraphics[angle=0, width=0.5\textwidth]{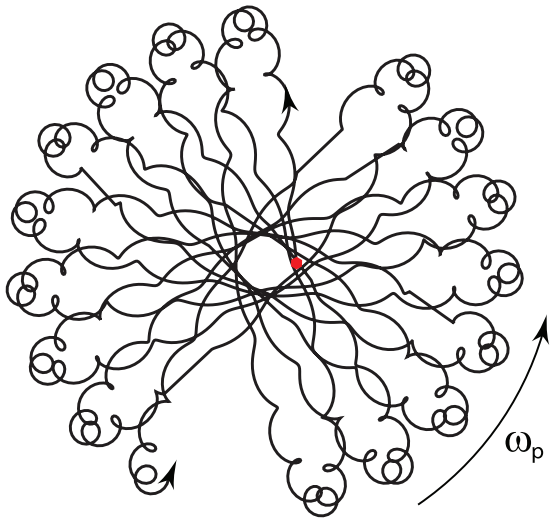}}
	\captionsetup{format=hang}
	\caption{  Prograde precession of a particle on a rotating saddle in the non-rotating frame for large angular velocity $\omega$ calculated according to equations \rf{eq:eqmotions}. In this illustration $\omega =20/9 \approx 2.2 $ (not even that large!).}
	\label{fig2}
\end{figure}

The particle trapped in the rotating saddle exhibits a prograde precession in the laboratory frame as illustrated in  Figure~\ref{fig2}. This means that the particle moves along an elongated trajectory that in itself slowly rotates in the laboratory frame with the angular velocity $\omega_p$ in the same sense as the saddle.  Up to now this precession has been explained by analyzing  explicit  solutions of the linearized equations (\cite{P1970,THB2002,HB2005}), leaving the underlying cause  of this precession somewhat mysterious.

 \subsection*{Deriving the equations of motion.} The motion of a point unit mass $ {\bf x} =(x,y) $ in any time--dependent potential $ U (x,y,t) $ is given by
\begin{equation}
	 \ddot {\bf x} =- \nabla U( {\bf x}, t)
	 \label{eq:eqmotion0}
\end{equation}
  (the gradient here is taken with respect to the $(x,y)$--coordinates).
The graph of our $ U( {\bf x}, t) $ is obtained by  rotating the graph of
$   U_0(x,y) =\frac{1}{2} (x ^2 - y^2) $, and we must
(i)   find the expression for this rotated potential  $  U ({\bf x},t) $, and (ii) compute $ \nabla U $ -- hopefully in an elegant way, without brute force.

The answer to (i) is simply
\begin{equation}
	U( {\bf x} ,t ) = U_0(R^{-1}   {\bf x}),
	\label{eq:rotated potential}
\end{equation}
where $ R= R(  \omega t)$ denotes the rotation through the angle $\omega t  $ around the origin (counterclockwise if $t> 0 $) and is given by the matrix
\begin{equation}
	R=R( \omega t ) = \left( \begin{array}{cc} \cos \omega t  & -\sin \omega t
	  \\ \sin \omega t   & \  \  \cos \omega t \end{array} \right) ; 	
	  \label{eq:R}
\end{equation}
Figure~\ref{fig:figrot} explains   why:
 \begin{figure}[thb]
	\center{  \includegraphics[angle=0, width=0.8\textwidth]{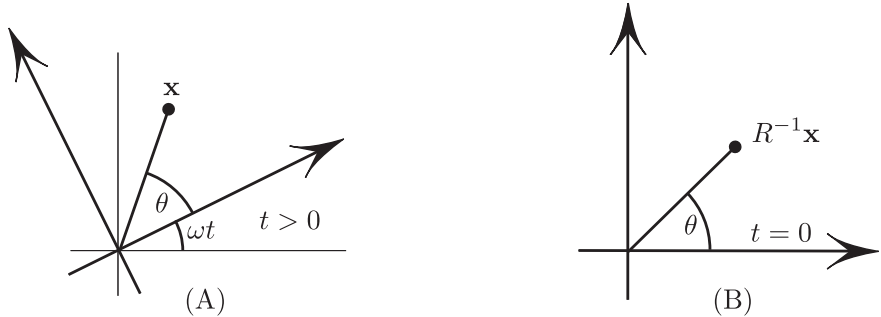}}
	\caption{Explanation of   (\ref{eq:rotated potential}).}
	\label{fig:figrot}
\end{figure}
Imagine looking down upon the graph of $ U( {\bf x} ,t ) $, Figure~\ref{fig:figrot}(A), focusing on a point ${\bf x}$;    let us  now rotate the graph {\it  together with} the point ${\bf x}$ clockwise by angle $ \omega t $, thereby turning the graph into the
 graph of $ U_0 $;  the result is shown in (B). Since the graph and the point ${\bf x}$ rotated together,  the height above ${\bf x} $ remained unchanged -- exactly what   (\ref{eq:rotated potential})  says!

(ii) to find $ \nabla U $ we could write      (\ref{eq:rotated potential}) in terms of  $x$ and $y$ and then differentiate,    but here is a more elegant  mess--avoiding way:
let us write $ U_0( {\bf x} )= \frac{1}{2} (x^2 - y ^2 ) $ as the dot product:
\be{neq14}
	U_0( {\bf x} ) =\frac{1}{2} (\rho  {\bf x}, {\bf x} ), \  \  \hbox{where}  \  \  \rho = {\rm diag}  (1,-1)
\ee
(note that $\rho$ is the mirror reflection in the $y$--axis), and use this in   (\ref{eq:rotated potential}):
\begin{equation}
	U( {\bf x} ,t ) = \frac{1}{2}(  \rho R^{-1}   {\bf x}, R ^{-1}  {\bf x} ) =
	\frac{1}{2}( R  \rho R^{-1} {\bf x}, {\bf x} ) =
	\frac{1}{2} ( S {\bf x} ,  {\bf x} );
		\label{eq:U}
\end{equation}
in  \rf{eq:U} we used the fact that  $R$ is an orthogonal matrix.
Multiplying out $ R \rho R ^{-1}  =S$ gives the matrix $S$ defined by equation \rf{eq:S}.

Now for any symmetric matrix $S$, one has $ \nabla ( S {\bf x} , {\bf x}) = 2 S{\bf x} $, as one can see with almost no computation\footnote{Indeed,   the definition of the gradient of a function $f$ states:  $ ( \nabla f( {\bf x} ), {\bf v})  \buildrel{def}\over{=} \frac{d}{ds} f( {\bf x}+ s {\bf v})_{s=0}$ for all vectors ${\bf v}$. In our case,  $ f( {\bf x} )  = ( S {\bf x} ,{\bf x}) $ and
$ \frac{d}{ds} f( {\bf x}+ s {\bf v})_{s=0}=  ( 2 S{\bf x} , {\bf v} )  $; comparing this with the left--hand side of the definition proves $  \nabla (S {\bf x} , {\bf x}) = 2 S{\bf x} $.}, and we conclude
\be{neq15}
	\nabla U ( {\bf x} , t)  =   S{\bf x};
\ee
Equation of motion   (\ref{eq:eqmotion0}) turns into   (\ref{eq:vectoreqmotions}), as claimed.

\paragraph {A ``spinning arrows'' interpretation of rotating saddle.}

\begin{figure}
	\center{  \includegraphics[angle=0, width=0.8\textwidth]{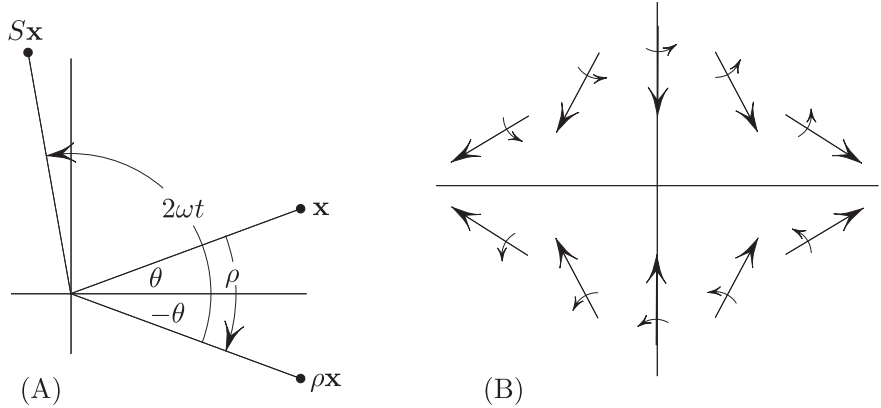}}
	\captionsetup{format=hang}
	\caption{ (A): $S$ is a reflection followed by a rotation. (B). Another interpretation of the rotating saddle: rotating the potential with angular velocity $\omega$ amounts to rotating each arrow of the vector field $ \langle x, -y\rangle $    with angular velocity $ 2 \omega $.}
	\label{fig1}
\end{figure}

Note that $S(\omega t) $ is a composition of the reflection with respect to the $x$--axis and the counterclockwise rotation through angle
$2\omega t$, Figure~\ref{fig1}(A). Therefore, for a {\it  fixed} ${\bf x}$ and increasing $t$, the vector $ S(\omega t) {\bf x}$
rotates counterclockwise with angular velocity $ 2\omega $. This leads to the following nice geometrical interpretation of the  governing equations   (\ref{eq:eqmotions})--(\ref{eq:vectoreqmotions}).
 The  force field  $ - S( \omega t) {\bf x} $ in our equations (\ref{eq:vectoreqmotions}) can be thought of in the following way, Figure~\ref{fig1}(B). Starting with  the stationary saddle vector field $ \langle x, -y \rangle $, we rotate each arrow of this field  with angular velocity $ 2 \omega $ counterclockwise; the result is our time--dependent vector field
 $ - \nabla U ( {\bf x} , t) = -  S( \omega t) {\bf x} $.

 \subsection* {Applications, connections to other systems.} Before getting to the point of this note,  we mention that equations   (\ref{eq:eqmotions})
  arise in numerous applications across many seemingly unrelated branches of classical and modern physics \cite{THB2002,S01,K2013a,K2013}; here is a partial list. They describe  stability of a mass mounted on a non-circular weightless rotating shaft subject to a constant axial compression force \cite{I1988,V95}, in plasma physics they appear in the modeling of a stellatron -- a high-current betatron with stellarator fields used for accelerating  electron beams in helical quadrupole magnetic fields \cite{P1970,S01,RMC1983}. In  quantum optics, equations  (\ref{eq:eqmotions})   originate in the theory of rotating radio-frequency ion traps \cite{HB2005}. In celestial mechanics the rotating saddle equations describe linear stability of the triangular Lagrange libration points $L_4$ and $L_5$ in the restricted circular three-body problem \cite{G1843,A1970,BB94}. In atomic physics the stable Lagrange points were produced in the Schr\"odinger-Lorentz atomic electron problem by applying a circularly polarized microwave field rotating in synchrony with an electron wave packet in a Rydberg atom \cite{BB94}. This has led to a first observation of a non--dispersing Bohr wave packet localized near the Lagrange point while circling the atomic nucleus indefinitely \cite{MGG2009}. Recently, the rotating saddle equations  (\ref{eq:eqmotions})   reappeared  in the study of confinement of massless Dirac particles, e.g. electrons in graphene \cite{N2013}. Interestingly, stability of a rotating flow of a  perfectly conducting ideal fluid in an azimuthal magnetic field possesses a mechanical analogy with the stability of Lagrange triangular equilibria and, consequently, with the gyroscopic stabilization on a rotating saddle \cite{OP1996}.

\subsection*{The result: a ``hodograph'' transformation}

  The main point of this note is to show that the rapid rotation of the saddle gives rise to an  unexpected Coriolis--like or magnetic--like  force   {\it  in the laboratory frame}; it is this force that is responsible for prograde precession.  To our knowledge this is the first example where   the Coriolis--like force arises in the inertial frame.

 To state the result we   assign, to any motion ${\bf x}= {\bf x} (t) $ satisfying   (\ref{eq:vectoreqmotions}),   its ``guiding center'', or ``hodograph''\footnote{The conventional hodograph transformation involves the derivatives of the unknown function. Hamilton referred to the path of the tip of the velocity vector of an orbiting planet as the hodograph (the word is a combination of the Greek words for {\it  path} and for {\it  trace, or describe}.).  Following Hamilton, in meteorology the hodograph is the trajectory of the tip of the velocity vector; in partial differential equations, Legendre's hodograph transformation
involves partial derivatives. We refer to \rf{eq:hodograph} as a hodograph transformation since $\bf u$ is a certain mixture of position and velocity.}
 \begin{equation}
	{\bf u} = {\bf x} - \frac{ \varepsilon ^2}{4} S( \omega t )( {\bf x}- \varepsilon J \dot {\bf x}  ),
	\label{eq:hodograph}
\end{equation}
where $  \varepsilon = \omega ^{-1} $  and where $J$ is the counterclockwise rotation by $ \pi /2 $:
\be{jeq}
	J = \left( \begin{array}{rr} 0  &  -1 \\  1 & 0 \end{array} \right) . 	
\ee

\begin{figure}[h]
	\center{  \includegraphics[angle=0, width=0.5\textwidth]{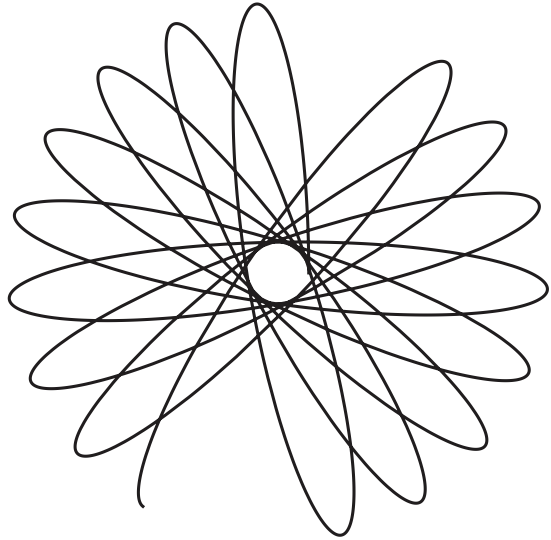}}
	\captionsetup{format=hang}
	\caption{  The motion of the ``guiding center'', governed by   (\ref{eq:truncated}), is the same as that of a Foucault pendulum. Here   $ \varepsilon = 0.45 $.}
	\label{fig3}
\end{figure}

We discovered that for large $\omega$, i.e. small $\varepsilon$,  the guiding center has a very simple dynamics:
\begin{equation}
	\ddot {\bf u} - \frac{\varepsilon ^3 }{4} J \dot {\bf u} +  \frac{\varepsilon ^2 }{4}    {\bf u} =O( \varepsilon ^4);
	\label{eq:u}
\end{equation}
that is, ignoring the $ \varepsilon ^4 $--terms, ${\bf u}$   behaves as a particle with a Hookean restoring force $ - \frac{\varepsilon ^2 }{4} {\bf u} $   and subject to the Coriolis-- or magnetic-- like force
$  \frac{\varepsilon ^3 }{4} J \dot {\bf u} $. Figure~\ref{fig3} shows a typical trajectory of the truncated equation
\begin{equation}
	\ddot {\bf u} - \frac{\varepsilon ^3 }{4} J \dot {\bf u} +  \frac{\varepsilon ^2 }{4}    {\bf u} =0.
	\label{eq:truncated}
\end{equation}
 This  is in fact the motion of a Foucault pendulum, and just as in the Foucault pendulum \cite{Khein1993}  the Coriolis--like term is responsible for prograde precession of ${\bf u}$ and thus of its ``follower''  ${\bf x}$.

 \paragraph {Angular velocity of precession in   (\ref{eq:truncated}) compared to earlier results.}
 Angular velocity of precession of solutions of the Foucault--type equation   (\ref{eq:truncated})  turns out to be $ \omega_p= \varepsilon ^3 /8 $. Indeed, writing the  equation in the frame rotating with
 angular velocity $ \omega_p$ gives rise to    a Coriolis term and a centrifugal term, and the system in that frame becomes
\be{neq1}
	\ddot {\bf z} + 2\omega_p J\dot {\bf z} - \frac{\varepsilon ^3 }{4} J \dot {\bf z} +  \frac{\varepsilon ^2 }{4}    {\bf z} + \omega_p ^2 {\bf z} =0.	
\ee
With the above value of $ \omega_p$, the second and the third terms cancel, and the resulting system
\be{neq2}
	 \ddot {\bf z} + (\varepsilon^2 /4+ \varepsilon ^6/64){\bf z} = 0
\ee
 exhibits no precession (all solutions are simply ellipses). We conclude that $ \  \varepsilon ^3 /8 $ is indeed the angular velocity of precession of ${\bf u}$. This simple expression fits with the earlier results and also gives the precession speed for the near--equilibrium motions of  Brouwer's particle, as we show next.

The equations of a rotating radio-frequency ion trap obtained in \cite{HB2005} reduce to
our equations \rf{eq:eqmotions} with $\varepsilon=\omega^{-1}=\sqrt{2q}$ after the re-scaling of time: $t=\tau\sqrt{2q}$, where $t$ is the dimensionless time in \rf{eq:eqmotions} and $\tau$ and $q$ represent here, respectively, the dimensionless time and a dimensionless parameter of the trap in \cite{HB2005}. Calculating the angle of precession $\omega_p t$, we find
\be{preca}
\omega_p t=\frac{\varepsilon^3}{8}t=\frac{\varepsilon^3}{8}\sqrt{2q}\tau=\frac{4q^2}{8}\tau=\frac{q^2}{2}\tau,
\ee
which yields the precession rate obtained by Hasegawa and Bollinger \cite{HB2005}:
\be{HB}
\omega_p^{HB}=\frac{q^2}{2}.\ee
Similarly, the dimensional precession frequency of the particle in Figure~\ref{fig:figsaddle} is: \be{hprs}
\Omega_p=\frac{g^2}{8r^2\Omega^3}.
\ee

\paragraph {A Coriolis--like force in the inertial frame.} If $  \frac{\varepsilon ^3 }{4} J \dot {\bf u} $ were a true Coriolis force, it would have been due to the rotation of the reference frame    with angular velocity $\frac{\varepsilon^3}{8}$ -- but our frame is inertial.  Alternatively, one can think of  $  \frac{\varepsilon ^3 }{4} J \dot {\bf u} $ as the Lorentz force  exerted on a  charge (of unit mass and of unit charge) in constant magnetic field $ B=\frac{\varepsilon^3}{4}$ perpendicular to the plane. Rapid rotation of the saddle gives
rise to a virtual pseudo--magnetic force (cf. \cite{BR1993,BS2010,BS2011}); as one implication, the asteroids in the vicinity of Lagrange triangular equilibria behave like charged
particles in a weak magnetic field, from the inertial observer's point of view.

\begin{figure}[h]
	\center{  \includegraphics[angle=0, width=0.5\textwidth]{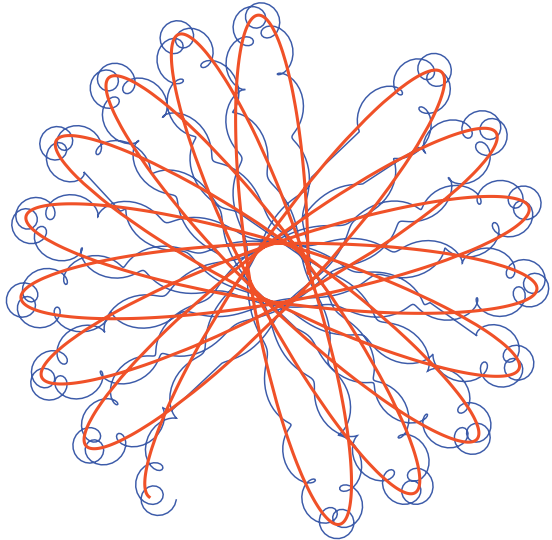}}
	\captionsetup{format=hang}
	\caption{A trajectory of the guiding center ${\bf u}$ governed by (\ref{eq:truncated}) (in red) tracking the corresponding trajectory ${\bf x}$ governed by (\ref{eq:vectoreqmotions}).  The view is in the inertial frame. $\varepsilon=0.45$.}
	\label{fig4}
\end{figure}


\paragraph {A numerical illustration.} Figure~\ref{fig4} shows a solution ${\bf x}$ alongside its ``guiding center'' ${\bf u}$. We note, as a side remark, that near the origin the solution follows the trajectory of the guiding center rather closely, reflecting the fact that  oscillatory micromotion is small near the origin, as is clear from the governing equations  (\ref{eq:eqmotions}).

\vskip 0.1 in
We discovered the  hodograph transformation (\ref{eq:hodograph})  via a somewhat lengthy normal form argument \cite{Levi1998,Levi1999} which, due to its length,
will be presented elsewhere.  Nevertheless, once the transformation (\ref{eq:hodograph}) has been produced, the statement (\ref{eq:u}) can be verified
directly by  substituting  (\ref{eq:hodograph})  into   (\ref{eq:vectoreqmotions}); we omit the routine but slightly lengthy details, but give a geometrical view of this transformation.

\begin{figure}[thb]
	\center{  \includegraphics[angle=0, width=0.5\textwidth]{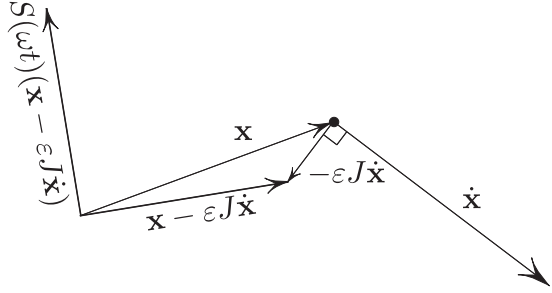}}
	\caption{A geometrical representation of the ``jiggle'' term $S(\omega t)({\bf x} -\varepsilon J \dot {\bf x})$ (without the $ \varepsilon ^2/4 $ factor) in the hodograph transformation. One should ``read'' the figure in the following order: $ x\mapsto \dot x \mapsto - \varepsilon J \dot x \mapsto x- \varepsilon J \dot x \mapsto S(x- \varepsilon J \dot x) $. A geometrical interpretation of $S$ is shown in Figure~\ref{fig1}(A).}
	\label{fig:jiggle}
\end{figure}

\paragraph{A geometrical view of the hodograph transformation.}
Our result says, in effect, that  the ``jiggle'' term
\be{neq5}
	 \frac{\varepsilon ^2 }{4} S(\omega t)( {\bf x} -\varepsilon J \dot {\bf x})
\ee
 in   (\ref{eq:hodograph}), when  subtracted from ${\bf x}$, leaves a smooth motion.\footnote{as far as the powers up to $\varepsilon ^3 $ are concerned.}
    Figure~\ref{fig:jiggle} gives a geometrical view of this term. It is still an open problem to find a simple heuristic explanation of choice of the term $ -\varepsilon J \dot {\bf x} $  in   (\ref{eq:hodograph}). Finding a heuristic explanation of the ``magnetic'' term $ \frac{\varepsilon ^3 }{4} J \dot {\bf u} $ remains an open problem as well.

\paragraph {Validity of the truncation.} Our result says that the fictitious particle ${\bf u}$, which shadows the solution ${\bf x}$, is subject to two forces
$ -(\varepsilon ^3/4) \dot {\bf u} - (\varepsilon ^2 /4){\bf u} $, plus a smaller $O(\varepsilon^4)$--force.

What is the cumulative effect of this force? One can show, using standard results of perturbation theory, that neglecting the $ O( \varepsilon ^4) $--term causes the  deviation less than  $  c_1 \varepsilon ^3 $
over the time $ | t | < c_2 \varepsilon ^{-2} $ for some constants $ c_1, \  c_2 $, for all $\varepsilon$   sufficiently small. As it often happens with rigorous results, this one is overly pessimistic: computer simulations show that ``sufficiently small'' is actually not that small: for example, $ \varepsilon = 0.45 $ in Figure~\ref{fig4}. In fact, the reason for such an unexpectedly good agreement is the fact that the error on the right--hand side is actually   $O ( \varepsilon ^6 )$ -- two orders better than claimed, as follows from an explicit computation by Michael Berry, \cite{B2015}. We do not focus on the analysis of higher powers of $\varepsilon$  because it would only add {\it  higher--order} corrections to the coefficients on the left--hand side of   (\ref{eq:u}), without affecting our main point (namely, that an unexpected Coriolis--like force appears).\footnote{According to M. Berry's computation \cite{B2015}  based on explicit solution of   (\ref{eq:eqmotions}), replacing  the coefficient  $ \varepsilon ^2 /4 $  in   (\ref{eq:truncated})  by
\be{neq6}
\frac{\varepsilon^2}{4}\left(1+\frac{3}{16}\varepsilon^4+\frac{11}{128}\varepsilon^8+\cdots\right)=-\frac{(1-\sqrt{1-\varepsilon^2})(1-\sqrt{1+\varepsilon^2})}{\varepsilon^2},
\ee
and replacing the coefficient $ \varepsilon ^3 /4 $ by
\be{neq7}
\frac{\varepsilon^3}{4}\left(1+\frac{5}{16}\varepsilon^4+\frac{21}{128}\varepsilon^8+\cdots \right)=\frac{2}{\varepsilon}-\frac{\sqrt{1-\varepsilon^2}+\sqrt{1+\varepsilon^2}}{\varepsilon}.
\ee
 yields the   equation for the exact guiding center (the expression for which is then given by   (\ref{eq:hodograph}) which includes higher order terms).
}

\begin{figure}[thb]
 \captionsetup{format=hang}
	\center{  \includegraphics{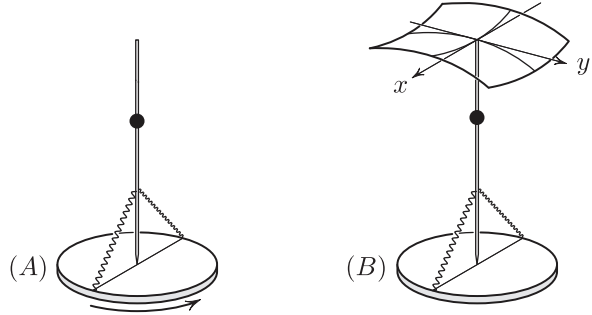}}
	\caption{A possible mechanical realization of the rotating saddle trap. Here $x$, $y$ are the angular variables,
	and the graph of the potential energy in terms of the angular variables $x$, $y$ is shown.}
	\label{fig:experiment}
\end{figure}

\subsection*{A proposed experiment.}
Figure~\ref{fig:experiment} illustrates a possible mechanical implementation of the rotating saddle trap (cf. \cite{I1988,V95}). As we had mentioned in the introduction, a ball rolling on the rotating saddle surface \cite{THB2002} is {\it  not } the right physical realization of the rotating saddle trap because (i)  the friction is very hard to eliminate, and, perhaps more importantly, because the rolling ball does not behave as a sliding particle. In fact, the rolling ball can be stable even  {\it on top}  of a sphere rotating around its vertical diameter \cite{W1997}!

A light rod with a massive ball  in Figure~\ref{fig:experiment}  is essentially an inverted spherical pendulum; the sharpened end of the rod, resting on the center of the platform, acts as a ball joint with the horizontal plane. The springs are attached to the rod.\footnote{Theoretically, we want to avoid transferring the rotation of the platform to  the rotation of the rod around its long axis (thereby affecting its dynamics), and thus the attachment should, theoretically, be via some frictionless sleeve. In practice, however, this frictionless sleeve will hardly make a difference:  the moment of inertia of the rod+ball around the long axis is negligible compared to the  moment  of inertia relative to a diameter of the platform, and thus the dynamics of the rod will be little affected by its axial rotation.} The height of the ball is adjustable, like in a \textit{metronome}. If the ball is placed sufficiently low then the two springs will stabilize the pendulum in the $x$--direction, Figure~\ref{fig:experiment}(B),\footnote{here $x$ and $y$ are angular variables.} and the potential acquires a saddle shape (since the $y$--direction is unstable):
\be{neq8}
	U_0(x,y) = \frac{1}{2}\ (a x^2 -  by ^2 ),
\ee
with $ a, \  b > 0 $,   ignoring higher powers of $x$, $y$. Here $ b = g/L $, where $L$ is the distance from the ball to the sharpened end of the rod. In the next paragraph we suggest a simple way  to adjust $L$ to make the two curvatures equal: $ a=b $, so that the linearized equations become
\be{neq9}
	\ddot {\bf x} + bS( \Omega t) {\bf x} = 0, \  \  b = \frac{g}{L}
\ee
a rescaled version of  (\ref{eq:eqmotions}). By rescaling the physical time $t$ to the dimensionless time $ \tau =  \sqrt{ a } t $, and introducing
\begin{equation}
	\omega = \Omega/  \sqrt{ a },
	\label{eq:oO}
\end{equation}
we transform the above equation into the dimensionless form   (\ref{eq:eqmotions}) (after renaming
$\tau$ back to $t$), which is stable if $\omega>1$ \cite{B1918,K2013,V95}.
 Now $ a=b=g/L $, where $L$ is the distance from the mass to the ball joint, and   (\ref{eq:oO})
gives us the length
\begin{equation}
	L= g \biggl(\frac{\omega}{\Omega} \biggl)  ^2.
\label{eq:L}
\end{equation}
The   $ 78 $ rpm of a vinyl record player corresponds to $ \Omega \approx 8.2 sec ^{-1}$, and the value  $ \omega \approx 2.2 $   referred to
 in Figure~\ref{fig2} corresponds to the height $ L\approx 71\,cm$. How short can we make the pendulum without losing stability? The cutoff length is $L\approx 14\,cm $, as follows from   (\ref{eq:L}) and the fact that   (\ref{eq:eqmotions}) is stable if and only if $ \omega > 1 $.  Incidentally, large $L$ corresponds to large $\omega$, according to (\ref{eq:L}) (for fixed rpm). This makes intuitive sense, since a ``natural'' unit of time in our system is the period $ 2 \pi  \sqrt{ L/g } $ of the oscillations along the stable axis of the stationary potential; for large $L$ this period is long, and during it the rotating potential will spin many times, corresponding to large $\omega$.

 \begin{figure}[thb]
 	\captionsetup{format=hang}
	\center{  \includegraphics{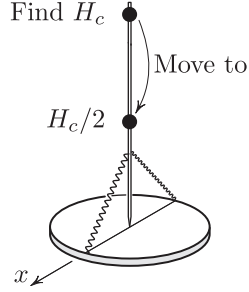}}
	\caption{How to find the length for which the curvatures of the saddle are equal and opposite:
	$ U_{xx}(0,0)=-U_{yy}(0,0) $: find by trial and error the height $ H_c $ at which the $x$--direction
	changes stability, and half $ H_c $.}
	\label{fig:experiment1}
\end{figure}

\paragraph {How to (easily) realize the saddle with equal principal curvatures.} It may seem (as it did to us initially) that one needs to measure the stiffnesses of the springs, the various lengths in Figure~\ref{fig:experiment}, the mass of the ball, and then use these to compute the value of $L$. Instead, here is a way to  avoid all this work. Referring to Figure~\ref{fig:experiment1}, {\it  adjust the position of  the massive ball  along the rod  to such critical height $H_c$ as to make the ball  neutrally stable in the $x$--direction:} if the ball is too high, it will be unstable in the $x$--direction; if the ball is too low,  it will be stable in the $x$--direction; a bisection method will quickly lead to a good approximation to  $H_c$. Remarkably,  the desired ``equal curvatures'' height of the ball is simply
 \be{neq10}
	 L = \frac{H_c}{2}.
\ee

\paragraph {An explanation.} The angle $y$ satisfies the   inverted pendulum equation
$ \ddot y - \frac{g}{L} \sin y = 0 $, which  for small angles is well approximated by
\be{neq11}
	\ddot y - \frac{g}{L} y = 0 .
\ee
Similarly, the linearized equation for the  angle $x$ is
\be{neq12}
	\ddot x + \biggl( - \frac{g}{L} + \frac{k}{L ^2} \biggl)   x = 0;
\ee
 $L$ is not yet chosen and $k$ depends on the parameters of the system, but not on $L$.\footnote{Specifically, $k$ depends on the springs' stiffness, on the angle they form with the diameter in Figure~\ref{fig:experiment} and on the ball's mass (to which it is inversely proportional). But an easier way to find $k$ is to note that $ k=gH_c$, as we show below.} Our goal is to find $L$ such that the coefficients in the above equations are equal and opposite, which amounts to asking for $L$ satisfying $ \frac{g}{L}= - \frac{g}{L} + \frac{k}{L ^2} $, or
\begin{equation}
	2Lg=k  .
	\label{eq:lk}
\end{equation}
 We now relate the unknown $k$ to $H_c$.  For the pendulum of the length $H_c$ the angle
$x$ satisfies $ \ddot x + (- \frac{g}{H_c} + \frac{k}{H_c ^2} ) x = 0 $, with  the coefficient $ - \frac{g}{H_c} + \frac{k}{H_c ^2}=0 $
since the equilibrium is neutral; this gives
\be{neq13}
	k = gH_c.
\ee
Substituting  $  H_c  = 2L $ into the last equation gives   (\ref{eq:lk}) -- precisely the condition for the equality of the coefficients in $U$.

\paragraph {Conclusion.} We showed that the rapid rotation of the saddle potential creates a weak Lorentz--like,  or a Coriolis--like force, in addition to an effective stabilizing potential -- all in the inertial frame. We also proposed a simple experiment to demonstrate the phenomenon.

 \vskip 0.3 in
\noindent  {\bf Acknowledgement} Mark Levi gratefully acknowledges support by the NSF grant   DMS-0605878.

 \end{document}